\documentclass[twocolumn,aps,prb,superscriptaddress]{revtex4}
\usepackage{graphics,bm}
\usepackage{amssymb}
\usepackage{amsmath}
\usepackage{physics}
\usepackage{epsfig}
\usepackage{epsf}
\usepackage{pdfpages}

\def\be{\begin{equation}} \def\ee{\end{equation}}
\def\beq{\begin{eqnarray}} \def\eeq{\end{eqnarray}}

\def\nn{\nonumber}

\begin{document}

\title{Real-Space Visualization of Quantum Phase Transition by Network Topology}
\author{Shehtab Zaman}
\affiliation{Department of Physics, Applied Physics, and Astronomy, Binghamton University - State University of New York, Binghamton, USA}

\author{Wei-Cheng Lee}
\email{wlee@binghamton.edu}
\affiliation{Department of Physics, Applied Physics, and Astronomy, Binghamton University - State University of New York, Binghamton, USA}

\date{\today}

\begin{abstract}
We demonstrate that with appropriate quantum correlation function, a real-space network model can be constructed to study the phase transitions in quantum systems.
	For the three-dimensional bosonic system, the single-particle density matrix is adopted to construct the adjacency matrix. 
We show that the Bose-Einstein condensate transition can be interpreted as the transition into a small-world network, which is accurately captured by the small-world coefficient. 
For the one-dimensional disordered system, using the electron diffusion operator to build the adjacency matrix,
we find that the Anderson localized states create many weakly-linked subgraphs, which significantly reduces the clustering 
coefficient and lengthens the shortest path. We show that the crossover from delocalized to localized regimes as a function of the disorder strength can be identified as 
the loss of global connection, which is revealed by the small-world coefficient as well as other independent measures like the robustness, the efficiency, and the algebraic connectivity. 
Our results suggest that the quantum phase transitions can be visualized in real space and characterized by the network analysis with suitable choices of quantum correlation functions.
\end{abstract}

\maketitle

\section{Introduction}
Complex network models have been employed to investigate various real-world systems such as social networks, information systems, biological systems, 
and physical systems \cite{girvan2002community, braiman1995taming}. 
Generally speaking, the architecture and dynamics of complex networks could be represented by graphs containing nodes and links.
A number of measures calculated from network models have been proposed to reveal the internal structures of graphs, 
which offers valuable insights for the real-world interacting systems of our interests.
In the pioneering work done by Watts and Strogatz,\cite{watts1998} the concept of the small-world network is introduced to describe systems whose nodes are highly clustered like 
regular lattices but have path lengths between nodes as small as random graphs. 
It has been shown that the small-world network has many advantages like fast information transmission, high synchronizability, etc., and many real-world systems such as human social groups, 
internet world in cyberspace, and biological systems have been shown to exhibit small-world 
charactersics\cite{montoya2002small,guimera2003self,jeong2001lethality,bullmore2009complex,bassett2008hierarchical,achard2007efficiency}.
It is remarkable that all these radically different systems could be described by the same network model, 
regardless the microscopic origins of the interactions that build up these systems.

Phase transitions in quantum systems have been one of the most important subjects in physics. 
By varying some physical parameters, the groundstate could change abruptly so that the system could exhibit very distinct physical properties.
In the field of the network science, it has been shown that some network models could be mapped into certain quantum systems, and consequently 
the nature of the network properties could be understood by the same ideas developed in quantum physics. 
For example, the model proposed by Bianconi and Barab\'asi\cite{bianconi2001} is shown to follow the Bose statistics and even undergo a Bose Einstein condensation.
The Fermi-Dirac statistics is shown to emerge in growing Cayley trees with thermal noise\cite{bianconi2002}, 
and a class of generalized statistics for networks has been proposed.\cite{garlaschelli2009}
While there is a growing interest in applying ideas from quantum systems to explore the relations in networks,\cite{zhu2000,lopez2005,sade2005,jahnke2008,bianconi2015,bianconi2015review,guo2016} 
the possibility of using the network science to explore the nature of phase transitions in quantum systems 
attracts much less attenion\cite{bianconi2012,halu2012,halu2013}. 
The key issue is the absence of an intuitive way to define nodes and links representing phase transitions in quantum systems.
Recently, Chou\cite{chou2014} has demonstrated that the topological phase transition in the $p$-wave superconductor can be characterized by the network science using the pairing 
amplitude to construct the adjacency matrix.
This analysis, however, is based on the mean-field Hamiltonian that assumes the existence of superconductivity, thus it can not be used to describe general cases of phase transition in quantum systems.

In this paper, we demonstrate that the phase transitions in quantum systems can be studied by network models with suitable quantum correlation functions defined in real space.
We test this idea in two cases that can be solved exactly.
For the three-dimensional non-interacting bosonic system, the single-particle density matrix is exploited to define the adjacency matrix for the weighted network.
We show that the Bose-Einstein condensate can be viewed as a small world in our network model due to the presence of the off-diagonal long-range order (ODLRO)\cite{yang1962}. 
For the one-dimensional disordered system, the weighted network is constructed using the electron diffusion operator.\cite{evers2008}
We find that the Anderson localized states create many weakly-linked subgraphs in our network model, which greatly reduce the clustering 
coefficient and lengthen the shortest path. We show that the crossover from delocalized to localized regimes with varying disorder strength can be reflected by 
the loss of global connection in the corresponding graphs, which can be clearly seen in the
small-world coefficient as well as other independent network measures like the robustness, the efficiency, and the algebraic connectivity. 
Our results suggest that the quantum phase transitions can be visualized in real space and characterized by the network topology with suitable quantum correlation functions.

\section{Bose-Einstein condensate in non-interacting three-dimensional bosonic system}
Consider a three-dimensional free boson system with fixed particle density $n$. The Hamiltonian can be written as
\be
H = \int d\vec{r} \hat{\psi}^\dagger(\vec{r})(\frac{\hbar^2\nabla^2}{2m}-\mu) \hat{\psi}(\vec{r}),
\ee
where $\hat{\psi}^\dagger(\vec{r})$ ($\hat{\psi}(\vec{r})$) creates (annihilates) a boson at position $\vec{r}$. Introducing the Fourier transformation of
$\hat{b}_{\vec{k}} = \frac{1}{\sqrt{V}}\int d\vec{r} e^{-i\vec{k}\cdot\vec{r}} \hat{\psi}(\vec{r})$,
we can diagonalize the Hamiltonian as
\be
H = \int \frac{d\vec{k}}{(2\pi)^3} (E(\vec{k}) - \mu) \hat{b}^\dagger_{\vec{k}}\hat{b}_{\vec{k}},
\ee
where $E(\vec{k}) = \frac{\hbar^2 k^2}{2m}$, and the particle density $n$ can be evaluated by
\beq
n&=&n_0 + n_e,\nn\\
n_e&=&\frac{1}{(2\pi)^3}\int \frac{d\vec{k}}{e^{\beta(E(\vec{k}) - \mu)} - 1}.
\eeq
$n_0$ is the particle density occupying the groundstate, while $n_e$ is the particle density occupying all other states.
The Bose-Einstein condenstate (BEC) is identified by $\mu$ being equal to the groundstate energy and the emergence of the off-diagonal long-range order (ODLRO) ($n_0 \neq 0$)\cite{yang1962},
which occurs at the temperature below the critical temperature $T_c$ of
\be
k_B T_c = 4\pi\frac{\hbar^2}{2m}\left[\frac{n}{\zeta(3/2)}\right]^{\frac{2}{3}}.
\ee
$\zeta(s)$ is the Riemann zeta function.

\begin{figure}
    \includegraphics[width=3.1in]{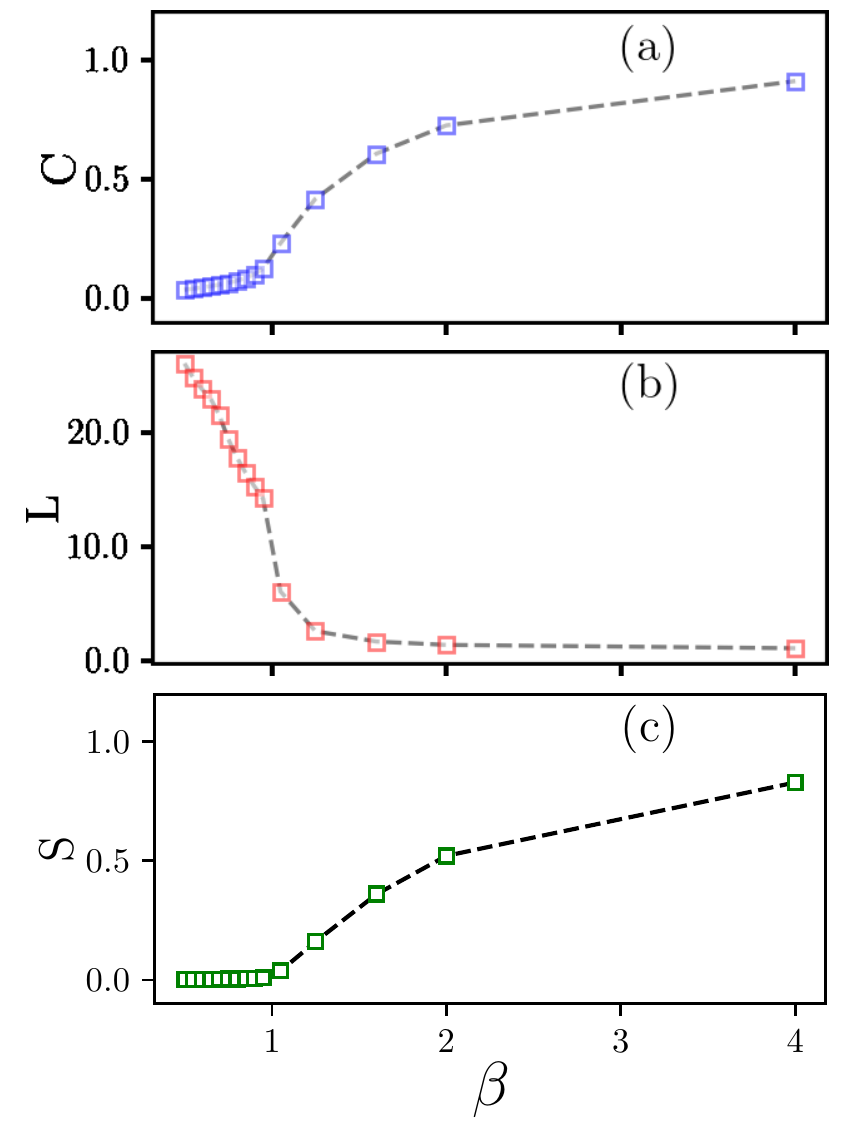}
	\caption{\label{fig:bec} (a)The average clustering coefficient, (b) average shortest path, and (c) small-world coefficient of the three-dimensional non-interacting boson system as a function
	of normalized Boltzmann factor defined as $\beta = 1/t, t= T/T_c$.}
\end{figure}

Our goal is to build a real-space network model to capture the BEC transition. A natural choice is 
the single-particle density matrix defined as
\be
\rho_{ij} \equiv \langle \hat{\psi}^\dagger(\vec{r}_i) \hat{\psi}(\vec{r}_j)\rangle = n_0 + f(\vec{r}_i -\vec{r}_j),
\label{odlro}
\ee
where
\be
f(\vec{r}_i -\vec{r}_j) \equiv \int d\vec{k} \frac{e^{-i\vec{k}\cdot(\vec{r}_i-\vec{r}_j)}}{e^{\beta(E(\vec{k}) - \mu)} - 1}.
\ee
Clearly, $f(\vec{r}_i-\vec{r}_j)$ goes to 0 as $\vert\vec{r}_i-\vec{r}_j\vert\to\infty$. 
In general, $\rho_{ij}$ descrbies the probability of a boson hopping from $\vec{r}_j$ to $\vec{r}_i$, which serves as a good 
quantity to represent the 'link' between different positions. 
Here we choose $N$ points along $x$ direction with equal spacing of $l_c = \frac{1}{\sqrt{4\pi}}\left(\frac{\zeta(3/2)}{n}\right)^{1/3}$. In other words, the position of $i$th point 
is $\vec{r}_i = (il_c,0,0)$, where $i=1,2,\cdots,N$.
We emphasize that the system is still three-dimensional, and we just choose points along $x$ direction.
Different choices of the set of points might give different values for the network measures, but the behavior as the function of temperature will be exactly the same.
Now we can construct a network model with these $N$ points being the nodes, and the adjacency matrix 
can be constructed based on the single-particle density matrix as
\beq
A_{ij} &=& 0, \,\,\,\,\,\,i=j,\nn\\
&=& \frac{\vert \rho_{ij}\vert}{Max(\vert \rho_{ij}\vert)}, i\neq j.
\label{aij}
\eeq
The graph built by Eq. \ref{aij} has weighted links bounded within $[0,1]$. For the network properties,
we have used $N=50$ in all the calculations presented below.
 
\begin{figure*}
	\includegraphics[width=6in]{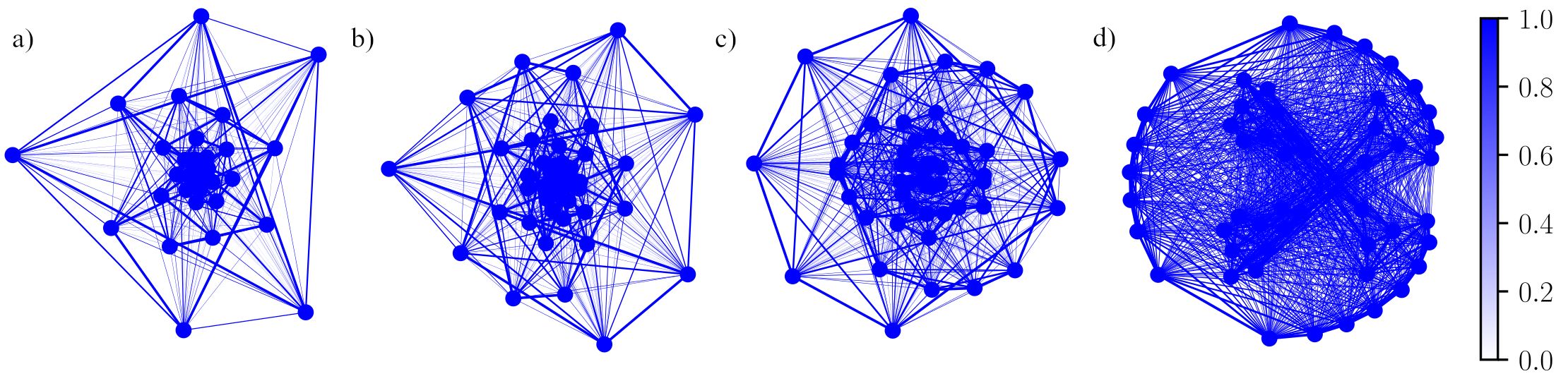}
	\caption{\label{fig:bec-aij} The force-directed graphs of the network model using the adjacency matrix defined in Eq. \ref{aij} for the three-dimensional non-interacting boson system as a function of normalized Boltzmann factor defined as $\beta = 1/t, t= T/T_c$ for (a) $ \beta = 0.85$, (b) $\beta=0.89$, (c) $\beta=0.94$, and $\beta =1.04$.}
\end{figure*}

We start from the local clustering coefficient that determines the connectivity of a given node in the graph. 
Following Ref. [\onlinecite{lopez2006}], we define the local clustering coefficient of $i$th node as
\begin{equation}
	C_{i} =\frac{1}{k_i (k_i -1)} \sum_{j,k \in \{N_i\}} A_{jk}
\label{ci}
\end{equation}
where $\{N_i\}$ is the set of neighboring nodes around $i$th node, and $ k_i$ is the number of nodes in $\{N_i\}$.
In an unweighted binary graph, $\{N_i\}$ is defined as the set of nodes $j$ with $A_{ij } = 1$. Since our graph is a complete graph with weighted links, 
$\{N_i\}$ should just be the set of all the nodes other than $i$, which means $k_i=N-1$ for every node.
Eq. \ref{ci} can naturally reproduce the clustering coefficient for an unweighted binary graph proposed in Ref. [\onlinecite{watts1998}]. 
It is noted that this definition emphasizes more on the connections between neighboring nodes around $i$th node without inputs from direct links to $i$th node, which 
is different from another popular definition\cite{onnela2005,chou2014} of
\be
C^O_{i} =\frac{1}{k_i (k_i -1)}\sum_{j,k \in \{N_i\}} \left(A_{ij}A_{jk}A_{ki}\right)^{1/3}.
\ee
$C^O_i$ counts the number of triangles attached to $i$th node with an effective intensity of $\left(A_{ij}A_{jk}A_{ki}\right)^{1/3}$, and consequently the direct links 
to $i$th node are as important as the links between its neighbors. Nevertheless, we find that general trends of the netowork properties are the same regardless the choice of 
the definition for the clustering coefficient.
To measure the clustering of the entire graph $G$, we can compute the average clustering coefficent as
\begin{equation}
	C(G) = \frac{1}{N} \sum_{i}C_i.
\end{equation}
Note that in our definition, $0\leq C(G)\leq 1$ and $C(G)=1$ only occurs in a complete graph with $A_{ij}=1$ for every $i\neq j$.

Another important quantity in the network science is the shortest path, which posts a subtle issue using the adjacency matrix $A_{ij}$ defined in Eq. \ref{aij}. 
In the network theory, if $A_{ij}$ is the weighted link between $i$th and $j$th nodes, 
its value is usually expected to be proportional to the 'distance' between these two nodes. This is, however, in contradition to the meaning of the single-particle density matrix which is 
proportional to the probability of a boson hopping from $\vec{r}_i$ to $\vec{r}_j$. 
In other words, in our graph, the larger $A_{ij}$ is, the shorter the 'distance' between $\vec{r}_i$ and $\vec{r}_j$ should be. 
As a result, we follow Ref. [\onlinecite{chou2014}] to use the 'inverse' of the adjacency matrix $A_{ij}$ to represent the 'direct' distance between nodes $i$ and $j$, denoted as 
$D_{ij} = 1/A_{ij}$. 
The shortest path between nodes $i$ and $j$, namely $d_{ij}$, is defined as the minimal length between $i$ and $j$ on the matrix of $D_{ij}$, which
can be found by Dijkstra shortest path algorithm.\cite{dijkstra1959} The global path length feature of the entire graph $G$ can be revealed by averaging the shortest path 
\begin{equation}
	L(G) = \frac{1}{N(N-1)} \sum_{i\neq j \in G} d_{ij}.
\end{equation}
Fig. \ref{fig:bec} presents $C(G)$ and $L(G)$ as a function of normalized Boltzmann factor $\beta = 1/t$ with $t=T/T_c$. 
It can be seen clearly that as the system enters the BEC region from the normal state ($\beta < 1$), $C(G)$ increases and $L(G)$ decreases abruptly at $\beta = 1$, 
exhibiting the BEC transition.
The kink at $\beta=1$ associated with the BEC transition can be further amplified in the small-world coefficient defined as 
\be
S(G) = \frac{C(G)}{L(G)},
\ee
which is shown in. Fig. \ref{fig:bec}(c).

These results can be well understood by the existence of the ODLRO given in Eq. \ref{odlro}.
In the normal state ($\beta > 1$), $n_0=0$ and $f(\vec{r}_i-\vec{r}_j)$ goes to zero as $\vert\vec{r}_i-\vec{r}_j\vert\to \infty$. Consequently, many elements in the adjacency matrix
are very small, and the normal state network tends to have a small $C(G)$ and long $L(G)$. 
In the BEC state ($\beta < 1$), every element in the adjacency matrix has the same order of magnitude due to
the ODLRO ($n_0\neq 0$), and the BEC network naturally has a large $C(G)$ and short $L(G)$ which are the key properties necessary for the small-world network.

To visualize the network constructed from Eq. \ref{aij}, we draw the force-directed graphs obtained by the Kamada-Kawai cost function
with the code developed by NetworkX\cite{networkx}. In the force-directed graphs, repulsive interactions similar to the electrical Coulomb interaction are introuced among every pair of nodes, and 
an attractive interaction between two nodes resembling the spring force is added with a spring constant depending on the value of $A_{ij}$. The balance of these forces determines the positions of 
nodes in a two-dimensional plane, which is shown in Fig. \ref{fig:bec-aij}.
The change of the graph topology can be seen directly as $\beta$ varies across the BEC transition. Compared to the graphs in the normal state, the distance between nodes 
are much more equal in BEC graph (Fig. \ref{fig:bec-aij}(d)), which is a clear signature for the small-world network.
Based on the above analysis, we conclude that the BEC transition corresponds to the transition in to a small-world network in the graph representation. 

\section{Anderson localization in 1D fermionic systems}
In this section, we employ the same idea to explore properties of the real-space network representing the disordered fermionic system.
For a proof of concept, we consider an one-dimensional periodic fermionic system with a Hamiltonian written as\cite{anderson1958}
\begin{equation}
	\hat{H} = - t\sum_{<i,j>}\hat{c}^\dagger_{i}\hat{c}_{j} + \sum_{i} \left[W\epsilon_{i}-\mu\right] \hat{c}^\dagger_{i}\hat{c}_{i}, 
	\label{dish}
\end{equation}
where $ \hat{c}^{\dagger_{i}}$ ($\hat{c}_{j}$) creates (annihilates) a fermion at site $i$, $t$ is the nearest-neighbor hopping integral, $W$ is the disorder strength, $\mu$ 
is the chemical potential, and $\epsilon_i$ is a random number bounded between $[-\frac{1}{2},\frac{1}{2}]$. We set $t=1$ in the following calculations.

To construct the adjacency matrix, we use the diffusion operator\cite{evers2008} defined as
\be
\Pi(i,j;\omega)\equiv \langle G^R_{ij}(\omega/2)G^A_{ji}(\omega/2)\rangle_{dis}.
\ee
Note that $\langle \cdots\rangle_{dis}$ denotes the average over different disorder configurations, and $G^{R,A}_{ij}(\omega)$ 
are the retarded and the advanced Green's functions between sites $i$ and $j$ 
\be
G^{R,A}_{ij}(\omega) = \langle i\vert \left[(\omega \pm i\eta)\hat{I} - \hat{H} \right]^{-1}\vert j\rangle.
\ee
$\Pi(i,j;\omega=0$) is a real and positive quantity which can be interpreted as the probability of an electron diffusing between sites $i$ and $j$ and is consequently 
a useful quantity to detect the degree of localization in a disordered system.\cite{evers2008}
As a result, we set up the adjacency matrix as
\be
\hat{A}_{ij} = \frac{\Pi(i,j;0)}{{\rm Max}(\Pi(i,j;0))}.
\label{dis-a}
\ee
By symmetry, $\hat{A}_{ij} = \hat{A}_{ji}$ and we have transformed the 1D disordered system to a network model with undirected but weighted links. 
We can therefore use the same definitions of the clustering coefficient, the shortest path, and the small-world coefficient introduced in the last section.
We solve the model Hamiltonian given in Eq. \ref{dish} with 500 sites. We fix the chemical potential to be $\mu=0$ and study the evolution of the graph as a function of the disorder 
strength $W$. For each disorder strength, we obtain $\Pi(i,j;0)$ by averaging over 5000 different disorder configurations.

\begin{figure}
    \includegraphics[width=3.1in]{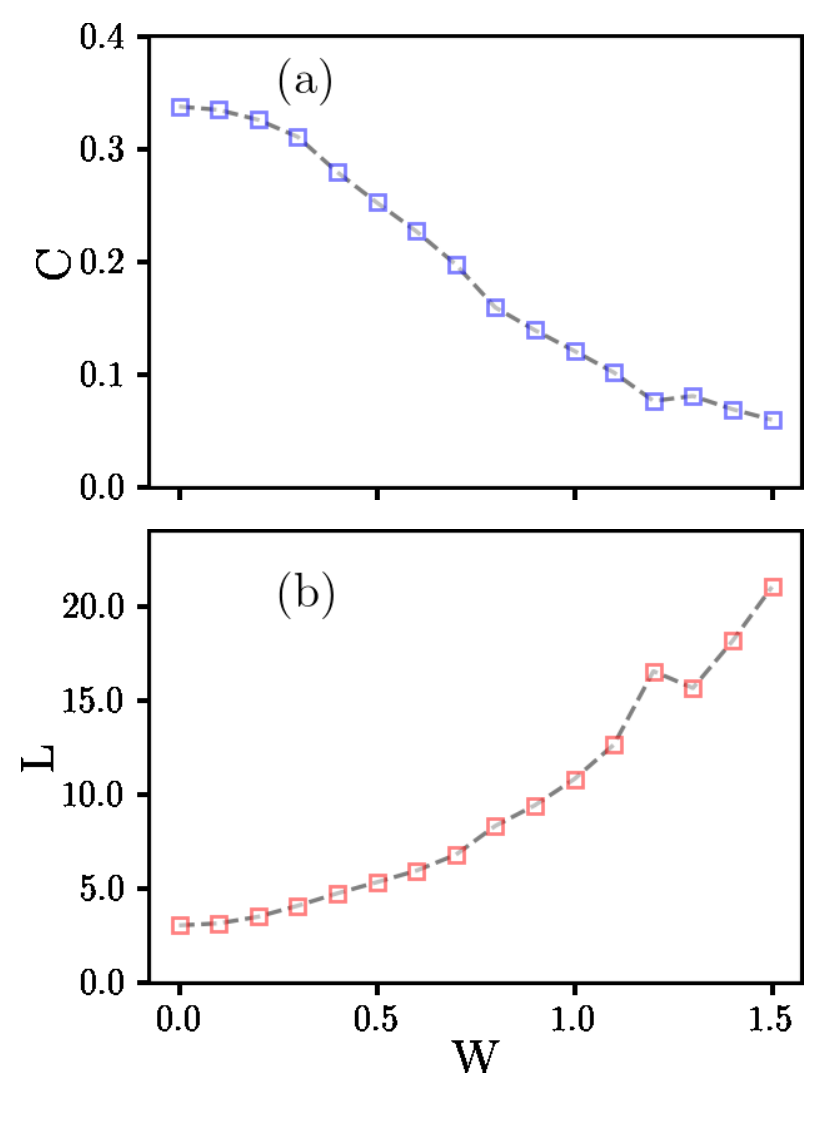}
	\caption{\label{fig:dis-cl} (a) The average clustering coefficient and (b) the average shortest path for the graph representing a 1D disordered system as a function of disorder strength $W$.}
\end{figure}

The average clustering coefficient and the average shortest path calculated based on the adjacnecy matrix in Eq. \ref{dis-a} are presented in Fig. \ref{fig:dis-cl} as a function 
of the disorder strength $W$.
Clearly, the clustering coefficient $C$ decreases as the disorder gets stronger, which can be understood by the nature of localization.
Because of the Anderson localized states, electrons are more likely trapped in localized states near particuar sites as the disorder strength gets stronger.
In the network model constructed from Eq. \ref{dis-a}, the reduction of the clustering coefficient with the increase of $W$ is attributed to many weakly-connected subgraphs resulted from
localized states. 
Moreover, since the probability of the electron hopping between different localized regions is generally small, the global average shortest path is expected to increase as a function of 
$W$, which is clearly shown in Fig. \ref{fig:dis-cl}(b).

\begin{figure}
    \includegraphics[width=3.1in]{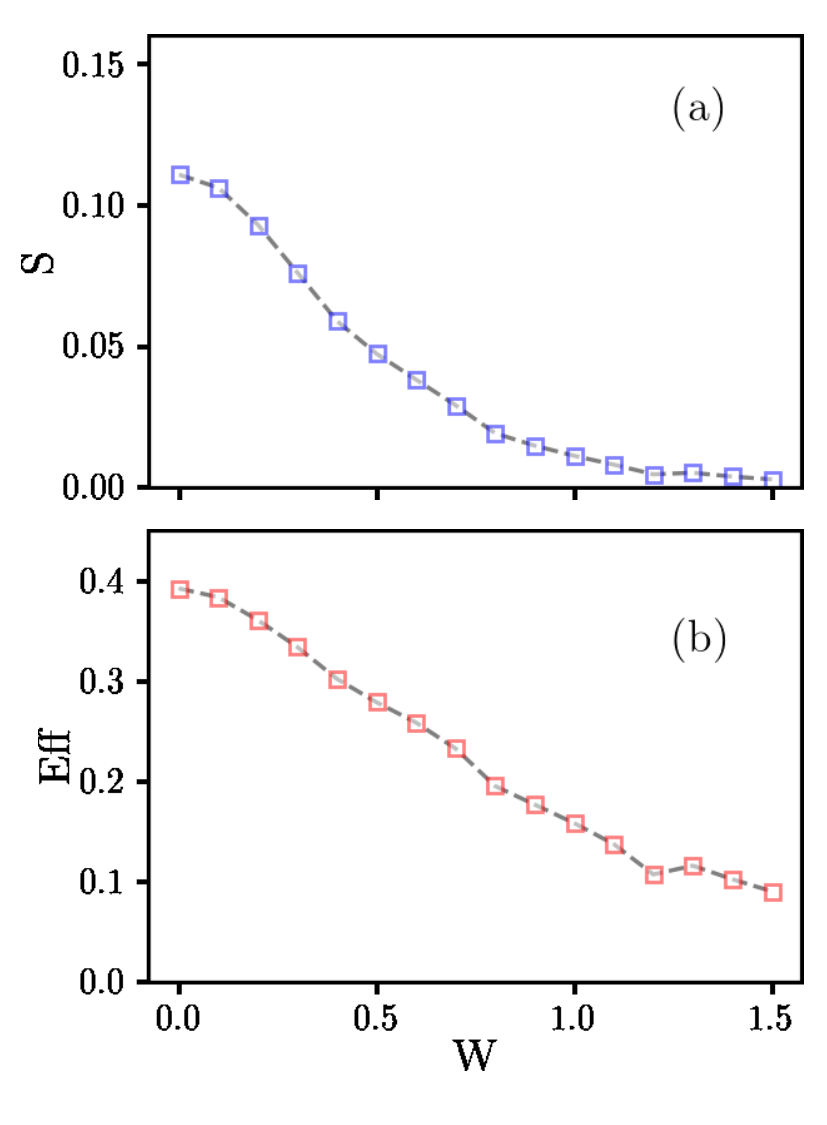}
	\caption{\label{fig:dis-se} (a) The small-world coefficient and (b) the efficiency for the graph representing a 1D disordered system as a function of disorder strength $W$.}
\end{figure}

We shall mention that because a one dimensional system is always in the localized regime in the presence of any disorder strength, we would not see a sharp transition as seen in the case of BEC. 
However, since we consider a finite-size system of 500 sites, we do see a crossover into a regime in which the localization length of every single-particle eigenstate is shorter than the system size.
The critical disorder strength ($W_c\sim 1.2 t$) for this crossover can be better characterized in the small-world coefficient plotted in Fig. \ref{fig:dis-se}(a).
An intriguing feature in Fig. \ref{fig:dis-se}(a) is that the small-world coefficient becomes almost zero as $W>W_c$. This feature indicates that the system is no longer globally connected
in the strong disorder limit, consistent with the fact that all the single-particle eigenstates have a localization length shorter than the system size, and consequently 
the electron diffusion is exponentially suppressed for any state.

\begin{figure}
    \includegraphics[width=3.1in]{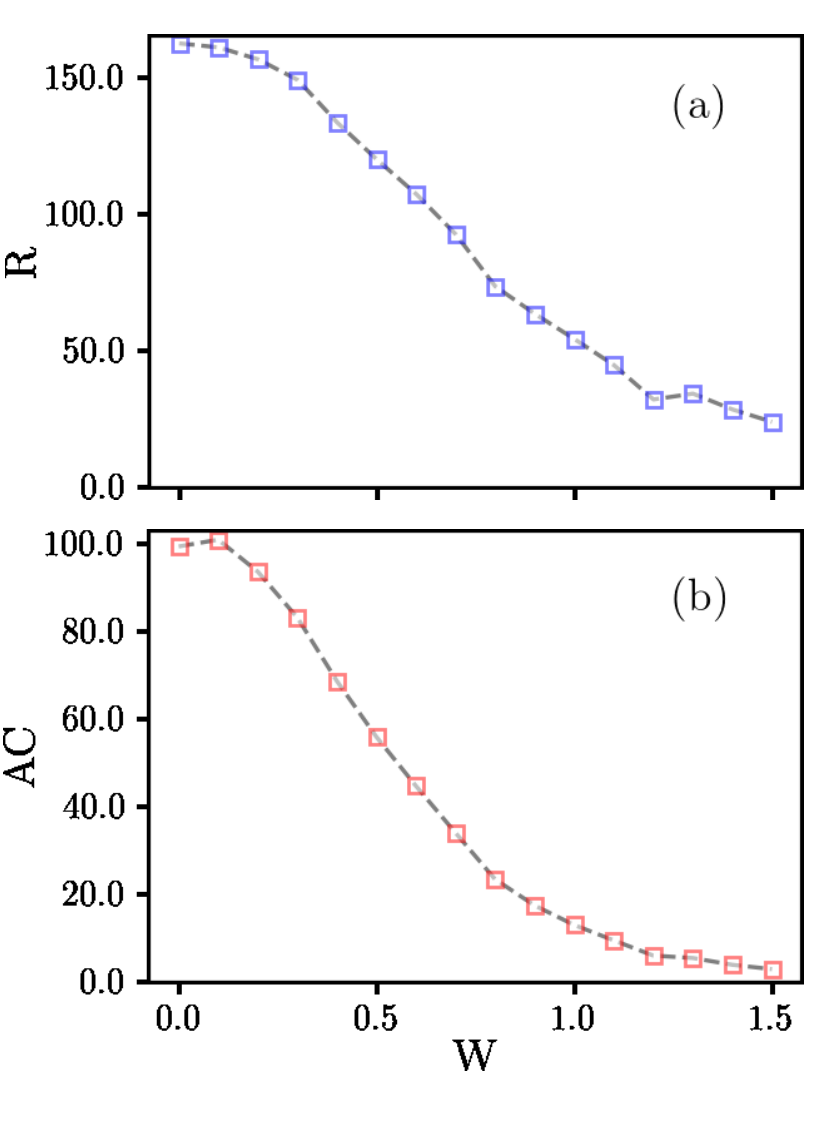}
	\caption{\label{fig:dis-rac} (a) The robustness $R$ and (b) the algebraic connectivity for the graph representing a 1D disordered system as a function of disorder strength $W$.}
\end{figure}

We can further study the loss of global connection by other network measures. 
For example, the reciprocal shortest distance matrix is used in calculating graph invariants such as the Harary Index\cite{harary1,harary2}, which has been employed to 
measure the compactness of molecular structures. A measure similar to the Harary index introduced in Ref. [\onlinecite{latora2001}]
has been exploited to indicate the efficiency of the graph, which is defined as the average of the inverse of the shortest path,
\begin{equation}
	{\rm Eff}(G) = \frac{1}{N(N-1)} \sum_{i\neq j \in G} \frac{1}{d_{ij}} 
\end{equation}
Higher ${\rm Eff}(G)$ corresponds to the increased ability of each node in the graph to disperse information simultaneously.
We plot ${\rm Eff}(G)$ as a function of $W$ in Fig. \ref{fig:dis-se}(b), which demonstrates the increased difficulty in transfering information between sites in the strong disorder regime.
Another quantity to analyze is the robustness $R$ defined as\cite{estrada1,estrada2,estrada3,estrada4,chou2014}
\begin{equation}
	R(G) = ln\left( \frac{1}{N} \sum_{i}e^{\lambda_i}\right),
\end{equation}
 where $\{\lambda_{i}\} $ is the set of eigenvalues of the adjacency matrix $\hat{A}_{ij}$. $R$ can be understood as a measure for the ability of the network to withstand nodes being removed 
without the global network features being affected. Generally speaking, a highly and globally connected graph tends to have a large value of $R$, 
meaning that removing a few nodes from the graph would not change the global properties of the graph significantly. The robustness $R$ as a function of $W$ is shown in Fig. \ref{fig:dis-rac}(a).
It can be seen that $R$ goes to zero in the strong disorder regime, which can again be attributed to the loss of the global connection.
In the strong disorder regime there are many weakly-linked subgraphs due to the localized states. 
As a result, removing a few nodes could disconnect completely the links between some subgraphs, and the connections between these affected subgraphs could be completely lost, 
which explains the vanishing value of $R$.

Lastly, we analyze the algebraic connectivity proposed by Fiedler\cite{fiedler1973}, which is another gauge of the connectedness of a graph\cite{wu2011spectral}. 
In order to obtain the algebraic connectivity, we introduce the degree matrix as
\beq
K_{ij}&=& 0, i\neq j,\nn\\
&=&\sum_k A_{ik}, i = j.
\eeq
The Laplacian matrix of the graph can therefore be constructed by
\begin{equation}
	\hat{L} = \hat{K} - \hat{A},
\end{equation}
It can be shown that the smallest eigenvalue of $\hat{L}$ is always zero for any graph, which corresponds to the eigenvector of $(1,1,1,\cdots)^T$. 
The second smallest eigenvalue, identified as the algebraic connectivity, is non-zero if and only if the graph is connected, and 
its value is proportional to the degree of the connectedness in a graph. \cite{de2007old,wu2011spectral}.
The algebraic connectivity for the graph representing the disorder system is plotted in Fig. \ref{fig:dis-rac}(b), which shows the same trends as the small-world coefficient, the robustness, and 
the efficiency.

\begin{figure}
    \includegraphics[width=3.1in]{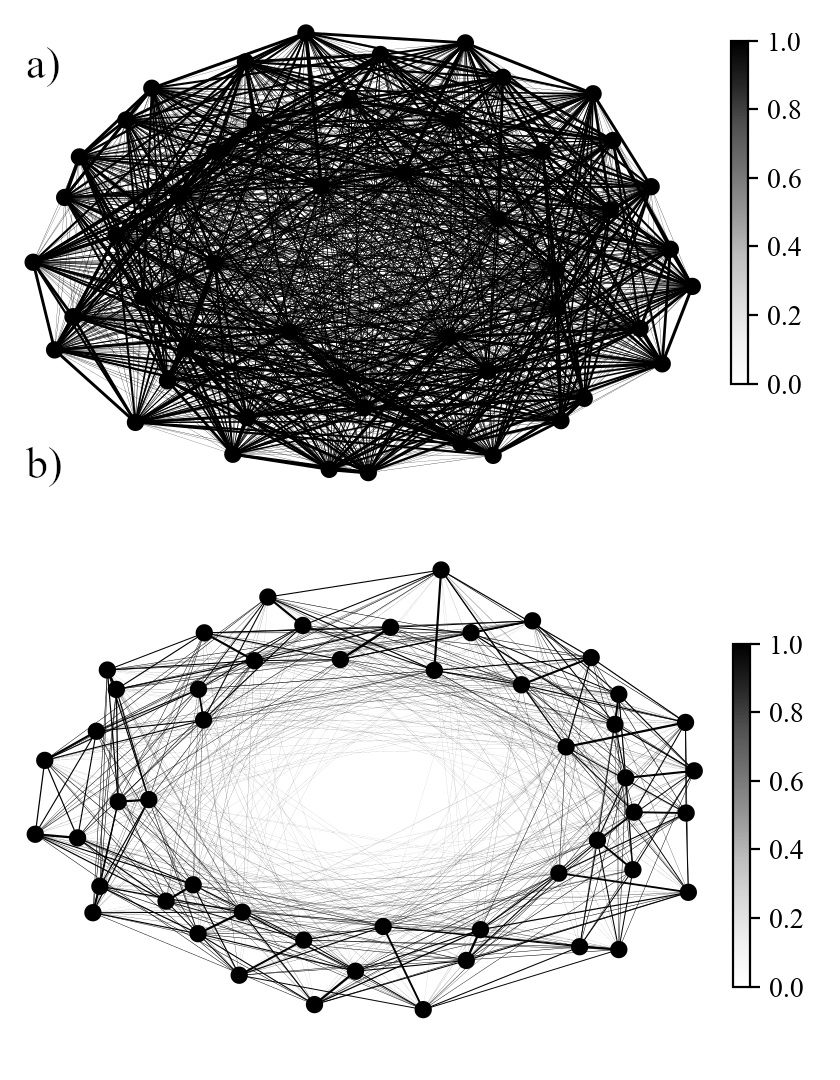}
	\caption{\label{fig:dis-aij} The force-directed graphs for the 1D disordered system in (a) delocalized ($W=0.5$) and (b) localized ($W=1.5$) regimes. For the best quality, we only plot 50 out of 500 nodes so that the distances between nodes can be clearly seen.}
\end{figure}

To directly visualize the subgraphs in our network, we draw the force-directed graphs\cite{networkx} in Fig. \ref{fig:dis-aij}. In the delocalized regime
(Fig. \ref{fig:dis-aij}(a) for $W=0.5$), nodes are more evenly distributed, indicating no tendency to form subgraphs.
On the other hand, in the localized regime nodes tend to stay closer to particular neighbors, and consequently a lot of subgraphs can be observed directly in the plot
(Fig. \ref{fig:dis-aij}(b) for $W=1.5$).

\section{Conclusions}
In this paper, we have studied two cases of phase transitions in quantum systems using the real-space network models. 
The key step is to choose a suitable quantum correlation function to construct the adjacency matrix, 
and the natural choice is the one encoding the 'probability' of quantum particles moving from one place to another in real-space. 
For the three-dimensional bosonic system we have chosen the single-particle density matrix, while for the one-dimensional disordered system we have chosen the 
electron diffusion operator.
We have shown that phase transitons in these model systems can be accurately captured by a number of network measures, which allows us to visualize the quantum phase transitions  
in real-space in terms of the network science. For example, the clustering coefficient measures the tendency of the neighbors around a node to connect, and the 
shortest path determines how fast the information can travel from one node to another. In the Bose Einstein condenstate (BEC), the off-diagonal long-range order provides global links between 
every node in the graph. As a result, the clustering coefficient increases and the shortest path decreases dramastically as the system undergoes the BEC transition, and in our 
network model the BEC transition can be characterized as a transition into a small-world network.
For the one-dimensional disordered system, becuase of the Anderson localized states, our network has many weakly-linked subgraphs that greatly reduce the clustering 
coefficient and lengthen the shortest path. We have shown that the crossover from delocalized to localized regimes as a function of the disorder strength can be well captured by the 
small-world coefficient as well as other independent measures like the robustness, the efficiency, and the algebraic connectivity.

Our formalism can be easily generalized to study the quantum phase transitions in other systems. For example, the quantum phase transitions in interacting fermionic systems can be analyzed by
network models constructed by two-particle density matrices\cite{yang1962} like spin-spin correlation function for magnetism, 
Cooper-pairing correlation function for superconductivity, and density-density correlation function for charge ordering.
Similar idea has been presented in Refs. [\onlinecite{valdez2017}] and [\onlinecite{sundar2018}] in which the quantum mutual information matrix constructed from one and two point correlators is 
introduced to study the quantum phase transitions in the Ising spin chain.
Our results demonstrate that the quantum phase transitions can be visualized in real space and characterized by the network topology with suitable quantum correlation functions.
 
\section{Acknowledgement}
We thank Christopher N. Singh and Matthew D. Redell for helpful discussions.
This work is supported by the Air Force Office of Scientific Research under award number FA9550-18-1-0024. 
Part of calculations used the Extreme Science and Engineering Discovery Environment (XSEDE) supported by National Science Foundation grant number ACI-1548562.

\end{document}